\documentclass[aps,prb,groupedaddress,twocolumn,showpacs]{revtex4}
\usepackage{graphicx}
\usepackage{multirow}

\begin{document}
\title{Critical exponents and fine-grid vortex model of  the dynamic vortex Mott transition in superconducting arrays.}

\author{Enzo Granato}

\address{Laborat\'orio Associado de Sensores e Materiais,
Instituto Nacional de Pesquisas Espaciais, 12227-010 S\~ao Jos\'e dos
Campos, SP, Brazil}

\begin{abstract}
We study the dynamic vortex Mott transition in two-dimensional superconducting arrays  in a magnetic field with $f$ flux quantum per plaquette. The transition is induced by external driving current and thermal fluctuations  near rational vortex densities 
set by the value of $f$, and has been observed experimentally  from the scaling behavior of the differential resistivity. Recently, numerical simulations
of interacting vortex models have demonstrated this behavior only near fractional  $f$. A fine-grid vortex model is introduced, which allows to consider both the cases of fractional and integer $f$.  The critical behavior is determined  from a scaling analysis of the current-voltage relation and voltage  correlations near the transition, and by Monte Carlo simulations.  The critical exponents for the transition near $f=1/2$ are consistent with the experimental observations and previous numerical results from a standard vortex model. The same scaling behavior is obtained for  $f=1$, in agreement with experiments. However, the estimated correlation-length exponent  indicates that even at integer $f$, the critical behavior is not of mean-field type.

\end{abstract}

\maketitle

\section{Introduction}

Superconducting arrays provide an interesting testing ground for equilibrium and nonequilibrium phase transitions. 
They can be realized as two-dimensional (2D) arrays of coupled superconducting regions or ``grains', with well controlled parameters,
being useful model systems of inhomogeneous superconductors, when phase fluctuations of the superconducting
order parameter play a major role \cite{newrock2000,benz88,geerligs,vdzant,ling1996,batur11,valles16}

Recently, remarkable nonequilibrium phase transitions induced by an applied current, have been revealed through experiments on a square array of  superconducting islands coupled by the proximity effect on a metallic film, in a perpendicular magnetic field \cite{poccia15,lankhorst2018}.   
The signature of the transition appears in the behavior of the  differential resistivity at low  temperatures, showing reversal of a minimum in to a maximum near certain values of the vortex density for increasing driving currents, and a corresponding scaling behavior as a function of current and vortex density near the transition. The transition has been identified as a classical analog of the dynamic quantum  Mott insulator transition \cite{nelsonvinokur93,tripathi2016,li2015,sankar2019}, with vortices playing the role of quantum particles.  Dynamic vortex Mott transitions were clearly identified near  integer vortex density $f=1$ and fractional vortex density $f=1/2$. The simplest model for such superconducting system consists of an ideal Josephson-junction array in an external magnetic field \cite{teiteljaya83,eg18} on the same lattice as the superconducting grains, where logarithmically interacting vortices are located at plaquette centers, which act as pinning sites. The average vortex density corresponds to the frustration parameter $f$, defined as the  number of flux quantum per plaquette.
%The equilibrium phase transitions of a JJA are strongly dependent \cite{teiteljaya83} on the value of $f$ and the geometry of the lattice \cite{shih1985,eg13}.  
The scaling behavior of the differential resistivity  observed experimentally \cite{poccia15,lankhorst2018} was found to be  described by a single critical exponent $\epsilon$.  This behavior has been demonstrated in recent numerical simulations of interacting vortex models \cite{rade17,eg18} only near fractional  $f$.  Outstandingly, for integer $f$, the value $\epsilon=2/3$ found experimentally agrees with a mean field description of the nonequilibrium dynamics obtained by mapping the dynamic vortex Mott  transition into a non-Hermitian quantum problem  \cite{tripathi2016}. Nevertheless, to characterize the critical behavior, the dynamic critical exponents $z$ and correlation-length exponent $\nu$ are also required. Near $f=1/2$, a different critical exponent $\epsilon=1/2$ was observed  \cite{poccia15}, which is not consistent with this mapping.  Numerical results \cite{eg18} for $f=1/2$, obtained from simulations of logarithmically interacting vortices, found an exponent $\epsilon$ consistent with the experimental observations and also obtained an estimate of the dynamic exponent $z \sim 2$ and correlation-length exponent $\nu \sim 1$.  These critical exponents clearly indicate that the dynamic vortex Mott transition at fractional vortex densities belong to a different universality class. To fully characterize the transition for integer $f$, it should, therefore, be  of interest to have  similar information on the value of these critical exponents but, so far, there are no estimates available from experiments or numerical simulations for this case.  

In this work, we study the dynamic vortex Mott transition in superconducting arrays using a lattice model of logarithmic interacting vortices, both at fractional and integer $f$. Because in the standard vortex model on a periodic lattice \cite{teiteljaya83,eg18}, the properties at  integer $f$ are equivalent to $f=0$, it does not allow the study of this dynamic transition at nonzero integer vortex densities. To circumvent this problem, a fine-grid vortex model is introduced, allowing us to consider both the cases of fractional and integer $f$ while still keeping the simplicity of the original model. The critical behavior is determined  from a scaling analysis of the current-voltage relation and voltage  correlation near the transition, and by Monte Carlo simulations. We find that, for $f=1/2$, the dynamic transition is accompanied by a structural transition of the sliding vortex lattice. The critical exponent $\epsilon$ for $f=1/2$ is consistent with the experimental observations  \cite{poccia15} and previous numerical results from the standard vortex model \cite{eg18}. The same scaling behavior of the differential resistivity is obtained for  $f=1$, in agreement with experiments \cite{poccia15,lankhorst2018}. From the scaling analysis we find $\epsilon=1/2 \nu$ and using the experimental results for $\epsilon$ we then conjecture the values $\nu=1$, $z=2$ for  $f=1/2$ and $\nu=3/4$, $z=7/3$ for $f=1$,  which are consistent with the numerical results within the errorbars. The results indicate that even for integer $f$ the dynamic vortex Mott transition is not of mean-field type and, therefore, fluctuations should be taken into account to fully describe the critical behavior.

\section{Model and simulation}

In the standard vortex model of a 2D superconducting array, vortices can only be located at the centers of plaquettes of the lattice formed by the superconducting grains, which act as  pinning sites. The grid of available sites correspond to  the dual lattice of the array. The vortex Hamiltonian is given by \cite{teiteljaya83} 
\begin{equation}
H_v = 2 \pi^2 E_o  \sum_{i,j} (n_i- f) G^\prime_{i,j} (n_j -f),
\label{cghamilt}
\end{equation}
where vortices are represented by integer charges $n_i$ ($n_i=0, \pm 1 ...$) at the sites  $  r_i=(x_i,y_i)$ of the dual lattice, constrained by the neutrality condition, $\sum_i (n_i -f )=0$.  $f$ is the number of flux quantum $\phi_0=hc/2e$ per plaquette of area $S$ introduced by the external magnetic field $B$, $f=BS/\phi_0$, and its value sets the average density of vortices. This vortex representation can  be obtained  following  a  standard  procedure \cite{jose1977} in which the usual Josephson-junction array model  \cite{teiteljaya83}, in terms of the phases of the local superconducting order parameter and Josephson coupling $E_0$, is  replaced by a periodic Gaussian model,
leading to explicit vortex variables  $n_i$. The vortex interaction is given by $G^\prime_{ij} = G( r_i - r_j) - G(0)$, where  $G( r)$ is the lattice 
Green's function  \cite{franz1995vortex,leeteitel94,hyman95}.   $G^\prime (\bf  r) $  
diverges logarithmically as  $ -\log(r)/(2 \pi) $ for large separations.  For a square lattice,
\begin{equation}
G({\bf  r}) =\frac{1}{L^2} \sum_k \frac{e^{i {\bf k}\cdot  {\bf r} }}{4-2 \cos({\bf k}\cdot {\bf a_1}) - 2 \cos({\bf k}\cdot {\bf a_2}) },
\end{equation}
where $L$ is the system size, $\bf  k$ are the reciprocal lattice vectors and $\bf a_1$, $\bf a_2$ are two perpendicular nearest-neighbor lattice vectors.   
When $f$ is an integer, a global change of the vortex charges $ n_i \rightarrow n_i +f$, shows that the properties of the model are the same as the case without external field, $f=0$. Because of this periodicity in $f$, the standard model does not discriminate
between zero and nonzero integer vortex densities, although it describes this transition for fractional vortex densities \cite{eg18}. 

\begin{figure}
\centering
\includegraphics[width=\columnwidth]{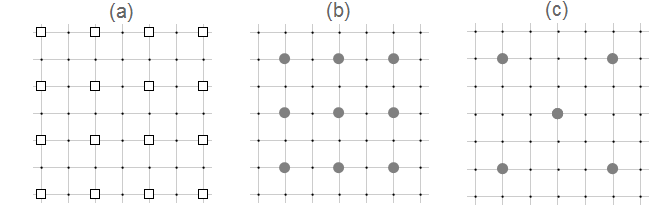}
\caption{(a) Schematic of the fine-grid vortex model. Squares represent superconducting grains and grid points the allowed positions for vortices; (b) and (c) represent the  vortex  configurations (filled circles) in the ground state for $f=1$ and $f=1/2$, respectively. }
\label{fgrid}
\end{figure}

In order to study both fractional and integer vortex densities within the same model, we introduce here a fine-grid vortex model on a square lattice.  
In addition to be located at the pinning sites of the array, vortices  can now also be located  at the junctions and at the grains of the array with a corresponding energy penalty (Fig. \ref{fgrid}). The spacing of the  grid of available sites is one half of the array spacing. 
The Hamiltonian of the fine-grid vortex model is given by
\begin{equation}
H_{fgv}  = 2 \pi^2 E_o  \sum_{i,j} (n_i- f^\prime) G^\prime_{i,j} (n_j -f^\prime)   +\sum_i E_i n_i^2 ,
\label{fghamilt}
\end{equation}
where the vortex charges $n_i$ are defined on the sites  $  r_i=(x_i,y_i)$  of the fine-grid lattice,  constrained by the neutrality condition, $\sum_i (n_i -f^\prime )=0$, where  $f^\prime=f/4$.  $E_i$  are additional vortex core energies: $E_i=E_J$ at the midpoint of the junctions between grains, $E_i=E_G>>E_J$ at the grains sites and $E_i=0$ at the pinning sites (plaquette centers). For sufficiently large $E_J$, the low-energy minimum  for $f=1$ (Fig. \ref{fgrid}b) corresponds to  a vortex configuration where there is  one  vortex  at each pinning site and for $f=1/2$ (Fig. \ref{fgrid}c), there is one vortex at alternating pinning sites. 

We  study  the  nonequilibrium  response  of  the superconducting array under an applied driving current
by  driven Monte Carlo (MC)  simulations \cite{leeteitel94,hyman95,gK98,eg98} of  the fine-grid vortex  model. 
The vortex dynamics is assumed to be overdamped. An external force is included, representing the effect of the driving current density $J$ on the vortices, acting as  a Lorentz force transverse to the velocity, leading to  an   additional   contribution   to   the   energy   in   Eq.  (\ref{fghamilt}),  $- (h/2 e) J  \sum_i n_i x_i $, 
when $J$ is  in the $\hat y$ direction.  The  MC  time  is
identified as the real time $t$ with the unit of time $dt=1$, corresponding to a complete MC pass through
the lattice. A MC step consists of adding a dipole of vortex charges to a nearest-neighbor charge pair $(n_i, n_j)$, 
using the Metropolis algorithm. Choosing this charge pair at random, the step consists of changing  $n_i \rightarrow n_i -1 $ and $n_j \rightarrow n_j +1 $, corresponding to the motion of a unit charge from $ r _i$ to $  r_j$. The move is accepted with
probability $ min[1,\exp(-\Delta H/kT) ] $, where  $\Delta H$ is the change in the energy. Periodic  boundary   conditions  
are used in systems of linear size $L$. The driving current $J$ biases the
added dipole, leading to a  net flow of vortices in the direction transverse to the current, if the vortices are mobile.  This vortex flow generates an electric field $E$ along the current which can be calculated (in arbitrary units) as $E(t)= \frac{1}{L} \sum_i \Delta Q_i(t) $,
%\begin{equation}
%E(t)= \frac{1}{L} \sum_i \Delta Q_i(t),
%\end{equation}
after each MC pass through the lattice, where  $\Delta Q_i=({\bf r_i} - {\bf r_j})\cdot {\bf \hat x} $ for an accepted vortex dipole excitation at the  sites $(i,j)$  and  $\Delta Q_i=0$  otherwise.  Due to the neutrality condition, $f^\prime$ is varied 
in multiples of $1/L^2$. Temperature $T$  is measured in units of $E_o/k_B$ and $J$ in units of $(2e/h) E_o$. 
%Most calculations were performed for system sizes ranging from  $L=12$ to  $48$.  
%Smaller systems were used to investigate finite-size effects.  

The results of the simulations presented in Sec. IV are for $E_J /E_0= 2$ with $E_G=4 E_J$. 
%Similar results were obtained for $E_J /E_0= 1$.
We use typically $ 5\times 10^5$ MC passes to compute time averages and the same number of passes to reach steady states. 

\section{Scaling analysis}

The expected behavior of the differential resistivity and other measurable quantities follows from  general arguments of the scaling theory
of a continuous dynamic transition occurring at a critical current  $J_c$.  Measurable quantities should scale with the diverging correlation length  $\xi \sim |\delta J|^{-\nu} $ and relaxation time $\tau \sim \xi^{z}$, where $ \delta J= J- J_c$, and  $\nu$ and $z$ are the correlation length and dynamic critical exponents, respectively. In particular, the differential resistivity scaling can be obtained in a similar manner as for 
the current-voltage scaling of inhomogeneous  superconductors \cite{fisher1991thermal,hyman95}, adapted to the present case of a transition at the critical current  $J_c $ and frustration $f_c$, with nonzero vortex density. 

\subsection{Current-voltage relation and differential resistivity}

Since the electric field $E$ generated by moving vortices with density $f$ and velocity $v$  is proportional to  $ f v$,  the singular contribution to $E$ should scale as $E \sim \xi^{1-z}$. 
Crossover effects due to a change  $\delta f = f - f_c$ should occur when  $ |\delta f |\xi^2 \approx 1$, corresponding to an additional vortex in a correlated area, revealing that $\delta f$ is a strongly relevant perturbation and should therefore appear in the scaling function in the combination $\delta J/|\delta f|^{1/2 \nu}$.   As a function of $\delta J$ and $\delta f$,  one then expect the current-voltage scaling  \cite{eg18}
\begin{equation}
E(J,f) = F_o(J,f) + |\delta f|^{\beta \epsilon} F_1(\delta J /|\delta f|^\epsilon),
\label{IVscal}
\end{equation}
where  $F_o$ is a regular contribution, analytic in $\delta J$ and $\delta f$, and $F_1(x)$ is a scaling function with $ F_1(0)=c$, a constant. The exponents $\beta$ and $\epsilon$ are determined by  the correlation length and dynamic critical exponents as 
\begin{eqnarray}  \label{expon}
\beta & = & (z-1 )\nu , \cr\
\epsilon & = & 1/2 \nu .
\end{eqnarray}
The scaling form for the differential resistivity  $dE/dJ$ can then be obtained from Eq. (\ref{IVscal}) as
\begin{equation}
\frac{dE(J,f)}{d J}- \frac{dE(J,f)}{d J}|_{J=J_c}= |\delta f|^{(\beta -1)\epsilon} H(\delta J /|\delta f|^\epsilon),
\label{diffscal0}
\end{equation}
with $H(0)=0$. We have neglected the $\delta J$ dependence of  $ dF_o(J,f)/d J $. 

The above scaling form reduces to the one used in the experiments \cite{poccia15,lankhorst2018} when $\beta=1$,
\begin{equation}
\frac{dE(J,f)}{d J}- \frac{dE(J,f)}{d J}|_{J=J_c}=  H(\delta J /|\delta f|^\epsilon),
\label{diffscal}
\end{equation}
which depends on a single exponent, $\epsilon$. The experimental data  for the differential resistivity is well described by this scaling form, both for fractional and integer frustration $f$. However, to fully characterize the critical behavior, the critical exponents $z$ and $\nu$ are also required. 

With $\beta=1$ and using the exponent relations in Eq. (\ref{expon}) we  can conjecture the values of the other exponents $\epsilon$, $ \nu$ and $z$, assuming one of them. The usual exponent for relaxation dynamics,  $z=2$, implies that $\nu=1$, which leads to a crossover exponent  $\epsilon=1/2 \nu =0.5$. Remarkably, this  value agrees with the experimental results \cite{poccia15} for $f =1/2$ and also with the numerical results from the standard vortex model \cite{eg18}. For $f=1$, however, the experiments find a different value \cite{poccia15,lankhorst2018}, $\epsilon=2/3$. Assuming this value for $\epsilon$, we get $\nu=3/4$ and   $z =7/3$. In the next Section, we compare these conjectured values with the numerical results obtained with the present fine-grid vortex model.

\subsection{Relaxation time}

To determine the  values of  the critical exponents $z$ and $\nu$ from numerical simulations, we performed a  scaling analysis of the relaxation time $\tau(J,f)$, obtained from the voltage time correlation function  
\begin{equation}
C(t)= \frac{<V(t) V(0)> - <V(t)>^2}{<V(t)^2>-<V(t)>^2} .
\label{tcorrel}
\end{equation}
Near the transition, $\tau$ can be estimated from the expected time dependence of $C(t)$ at long times, 
$C(t)\propto e^{-t/\tau}$.
Since $\epsilon=1/2 \nu$,  the relaxation time   $\tau \sim \xi^{z}$ should then satisfy the  scaling form 
\begin{equation}
\tau  |\delta f|^{z /2} = G(\delta J /|\delta f|^{1/2 \nu}) ,
\end{equation}
in the absence of finite-size effects. 
The critical  exponents $\nu$ and $z$ can be estimated from the  best data collapse in a plot of  $\tau  |\delta f|^{z /2}$ versus $\delta J /|\delta f|^{1/2 \nu}$, satisfying this scaling form.  To minimize the finite-size effects, this data collapse is performed for large systems and in a range of $f$ not too close to $f_c$. 

However, when the correlation length $\xi$ becomes comparable to the system size $L$, the scaling function will also depend on the dimensionless ratio $L/\xi$ as
\begin{equation}
\tau  |\delta f|^{z /2} =  F_2(\delta J /|\delta f|^{1/2 \nu},\delta J L^{1/\nu}).
\label{tauscal}
\end{equation}
This makes the numerical determination of the critical parameters very complicated due to the presence of  two scaling variables. As a simplification, in this case we consider data at current densities  and system sizes such that $\delta J L^{1/\nu} $  is equal to a constant value. Then, the scaling function $F_2$ depends only on a single variable $\delta J /|\delta f|^{1/2 \nu}$.

At the transition, $J=J_c$, the correlation length is cutoff by the system size $L$ and the relaxation time $\tau$ should satisfy the finite-size scaling form
\begin{equation}
\tau/ L^z  = F_3(L^2|\delta f|) .
\label{tauscalb}
\end{equation}
%Indeed, as shown in Figs.  \ref{tau}a and \ref{tau}b, a reasonable data collapse  according to the above  scaling forms  is obtained with $z\approx 2$ and $%\nu\approx 1$.
%, near the estimate of  $J_c$ from Fig. \ref{ivf12}. 

\subsection{Vortex correlation}

%The critical exponent $\nu$ was also determined from the scaling behavior of the vortex correlation. 
The dynamic vortex Mott transition should correlate with a change in the structure of the sliding state of the vortex lattice. 
For $f=1/2$, we find that this change can be quantified from the behavior of the structure factor  $S({\bf  k})$, which is 
% As can be seen from the snapshots of
%the vortex configuration in Fig. \ref{sf}a and Fig.  \ref{sf}b, obtained below and above the critical current $J_c$, respectively, the vortex lattice remains essentially ordered below $J_c$ but has a large number of defects above $J_c$. 
%The transition can then be determined from the
%bebavior of the structure factor $S({\bf  k})$, which is
a measure of the vortex correlations, 
\begin{equation}
S({\bf k})= \frac{1}{L^2} <|n({\bf  k})|^2>,
\label{dsf}
\end{equation}
where $n({\bf k})$ is the Fourier transform of the vortex variables $n_i$. The sliding ordered state below $J_c$  corresponds to sharp peaks in the structure factor $S({\bf  k})$  at the wave vectors ${\bf  k_o}$ of a periodic vortex structure.  Above $J_c$, the peaks broaden and become very small corresponding to a disorder phase. 
%The scaling behavior of $S({\bf  k})$ should be determined by the same critical exponent $\nu$ of the differential resistivity transition obtained above. 
Assuming a structural phase transition, $S({\bf  k_o})$ should then  satisfy the scaling form
\begin{equation}
S({\bf  k_o})  |\delta f|^{(1-\eta/2 )} = F_4(\delta J /|\delta f|^{1/2 \nu},\delta J L^{1/\nu}) ,
\label{sfscal}
\end{equation}
where $\eta$ is an additional critical exponent characterizing the power-law decay of vortex correlations at the transition 
$ <n_j n_l \ e^{i {\bf k_0} \cdot ({\bf r_j} - {\bf r_l}) }>\sim |{\bf  r_j} -{\bf r_l}|^{-\eta}$. At the transition, it should satisfy the scaling form 
\begin{equation}
S({\bf  k_o})/  L^{(2-\eta )} = F_5(L^2|\delta f|) .
\label{sfscalb}
\end{equation}
%As shown in Figs.  \ref{sfscal}a and \ref{sfscal}b, a reasonable data collapse  according to the above  scaling forms  is obtained with $\nu\approx 1$  and
%$\eta \approx 0.3$.

We also consider the scaling behavior of the finite-size vortex correlation length, which can be obtained from $S(k)$ as 
\begin{equation}
\xi(J,f) = \frac{1}{2\sin(k_m/2)}[S(k_0)/S(k_1) - 1]^{1/2}.
\label{cdef}
\end{equation}
Here  ${\bf k_1}={\bf k_0}+{\bf k_m}$ and ${\bf k_m}$  is the smallest nonzero wave vector of the lattice. This expression is a generalization of the usual second-moment correlation length \cite{amit2005} for an order parameter with nonzero wave vector $\bf k_o$.  Above the transition,  $J > J_c $, this definition  corresponds to  a finite-difference approximation to the infinite system correlation length
$\xi^2= -\frac{1}{S({\bf k})} \frac{\partial S({\bf k}) }{\partial ({\bf k}-{\bf k_0})^2} |_{{\bf k}={\bf k_0}}$,
taking into account the lattice periodicity. $\xi $  should satisfy the scaling form 
\begin{equation}
\xi  |\delta f|^{1/2} = F_6(\delta J /|\delta f|^{1/2 \nu},\delta J L^{1/\nu}) ,
\label{cscal}
\end{equation}
near the transition. At the transition, it should satisfy the scaling form
\begin{equation}
\xi/ L  = F_7(L^2|\delta f|),
\label{cscalb}
\end{equation}
which, interestingly enough, does not depend on the critical exponents.
To verify the scaling behavior of  Eqs. (\ref{sfscal}) and (\ref{cscal}), we use the same procedure as for  Eq. (\ref{tauscal}). 

\begin{figure}
\centering
\includegraphics[width=\columnwidth]{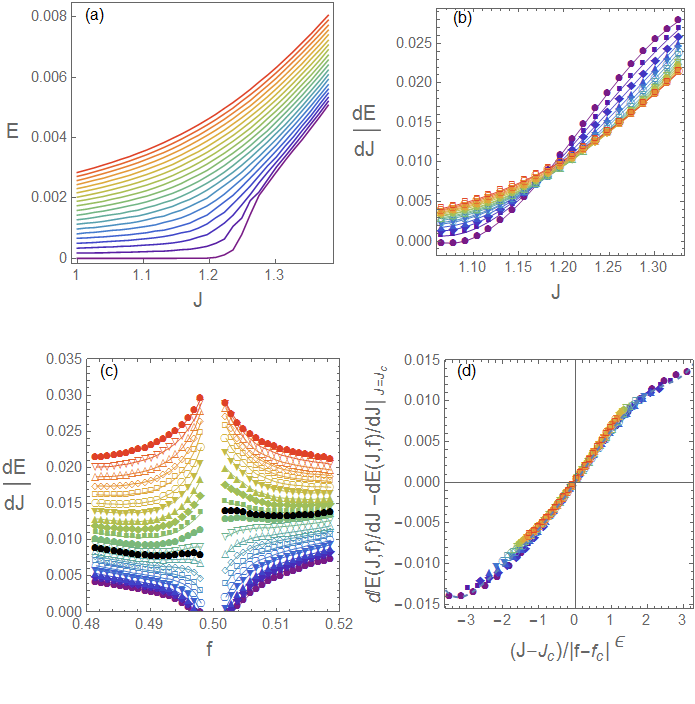}
\caption{(a) Nonlinear current-voltage characteristics ($J=I/ L$,  $E=V/L$ ) for increasing frustration $f$ near $f=1/2$. Temperature $T=0.2$ and system size $L=64$. From the bottom up,  $f$ increases from $0.5$ to $0.518555$ in $19$ equal steps. 
(b) Differential resistivity $\frac{d E}{d J}$ for $f \ge 0.50293$  near the transition. 
(c) $\frac{d E }{d J}$  as a function of $f$  for different $J$. From the bottom up, $J$ increases from $1.06552 $ to $1.32759 $ in $20$ equal steps. The separatrix $ dE/dJ_{J=J_c} $ is indicated by black dots.
(d) Scaling plot of $\frac{d E}{dJ}$ for $ f > 1/2$, with  $J_c=1.203$, $ f_c=0.5$ and $\epsilon=0.55$.}
\label{ivf12}
\end{figure}

\begin{figure}
\centering
\includegraphics[width=\columnwidth]{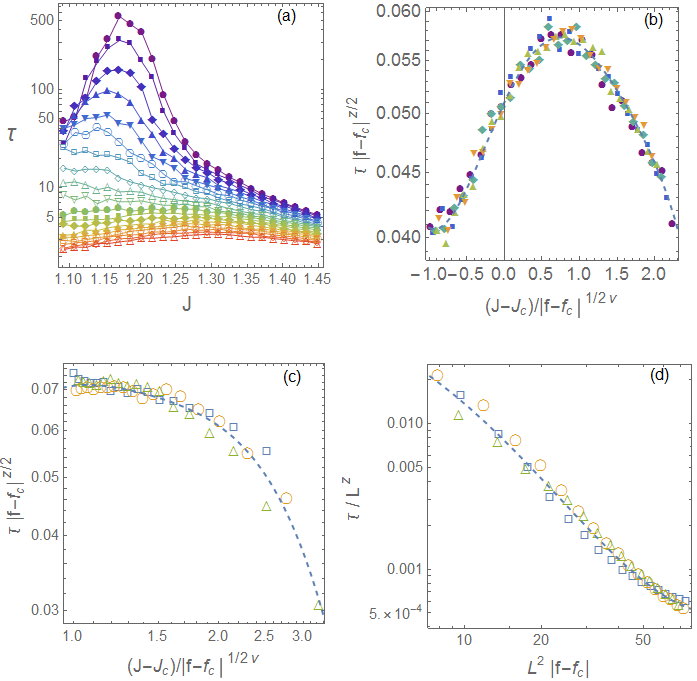}
\caption{ (a) Relaxation time  $\tau(J,f) $  at different values of $f$, near the dynamical transition for $f = 1/2$.  Temperature $T=0.2$ and system size $L=112$.  $f$ increases from $0.500957$ to   $0.518176$ in $18$ equal steps;
b)  Scaling plot for $f \ge 0.514349$ neglecting finite-size effects with $J_c=1.19$, $ f_c=0.501$, $\nu=1$ and $z=2$;
(c) Scaling plot with  $\delta J L^{1/\nu}=c $, a constant ($c= 7.09$),   with  $J_c=1.16$, $ f_c=0.501$,  $\nu=1.1$ and $z=1.9$. Open squares correspond to $L=48$, circles to $L=64$ and triangles to $L=80$;  (d) Scaling plot at $J_c=1.24$ with $z=2.1$. }
\label{tauf12}
\end{figure}

\section{Numerical simulations and discussion}

\subsection{Fractional vortex density: $f=1/2$ }

First, we consider the dynamic vortex Mott transition near  $f=1/2$ and check if the present fine-grid vortex model gives the same results as the previous simulations \cite{eg18} with the standard vortex model.  New results are also obtained from calculations of the vortex structure factor and correlation length. Figure \ref{ivf12}a shows the effect of increasing the frustration $f$ from
$f=1/2$ on the current-voltage (I-V) relation, in terms of the current density $J=I/L$ and the electric field $E=V/L$ at a temperature $T=0.2$. This temperature is well below the critical temperature of the equilibrium resistive transition, $T_c \sim 0.8$, which occurs at $J=0$.  A small increment in $f$ from $f=1/2$ leads to an increase in the slope of the current-voltage curve, which changes sharply at a critical value $J_c\approx 1.2$. Further increase of $f$ tends to  smooth out the slope of these curves near $J_c$.  This change of slope near $J_c$ can been seen much clearer in the behavior of the differential resistivity,  $dE/dJ$,  shown in  Fig. \ref{ivf12}b. To obtain smooth curves for $dE/dJ$ by numerical differentiation,  the current-voltage data in the small interval near $J_c$ was fitted to a low order polynomial.  Fig. \ref{ivf12}b reveals that the curves $dE/dJ \times J  $ for different $f > 0.502$ cross approximately at the same point  $J_c$. The crossing point is a manifestation of the underlying dynamic transition, with  $dE/dJ$ behaving approximately  as  a scaling invariant quantity and $f - f_c$ acting as a relevant perturbation \cite{eg18}, consistent with the scaling form of Eq. (\ref{diffscal0}) when $\beta=1$.  When the differential resistivity is plotted as a function of $f$ for different currents in Fig. \ref{ivf12}c, where data for $f < 1/2$ is also included,  there is a characteristic reversal of a minimum into a maximum near $f=1/2$ for increasing current density,  at  $J_c$. The main difference of this behavior in the present model and the standard model \cite{eg18} is the assymetry of the curves with respect to $f - 1/2$.  In  Fig. \ref{ivf12}d, we plot the data near the transition according to the scaling form  of Eq. (\ref{diffscal}), originally proposed in the experiments \cite{poccia15}.  Data for different $J$ and $f$  collapse into the same smooth curve when $J_c$, $f_c$ and $\epsilon $ have the appropriate values. The value obtained for the critical exponent,  $\epsilon = 0.55(7)$, is consistent with the one obtained from previous numerical simulations with the standard model \cite{eg18} and also with the experiments \cite{poccia15}, supporting the universality of this dynamical transition.  The reversal of the minimum in to a maximum and the data collapse characterized by a single exponent $\epsilon$ are the signatures of the dynamic vortex Mott  transition as observed in the experiments \cite{poccia15,lankhorst2018}. 

\begin{figure}
\centering
\includegraphics[width=\columnwidth]{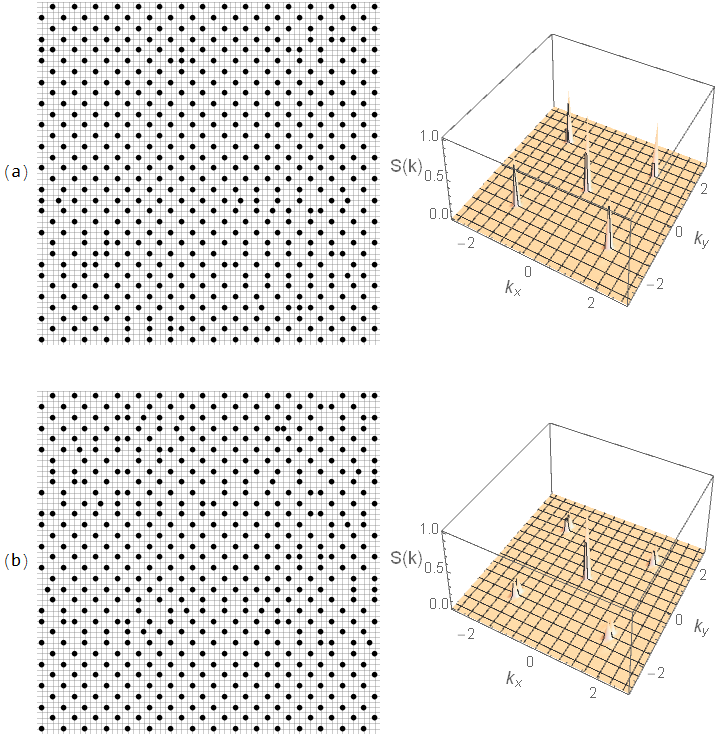}
\caption{ Snapshots of the vortex configurations  and  corresponding structure factor $S(\vec k)$  at   $J=1.17034 $ (a) and    $J=1.22276$ (b), below and above the dynamical transition, respectively, for $f = 0.505859$. }
\label{sf}
\end{figure}

\begin{figure}
\centering
\includegraphics[width=\columnwidth]{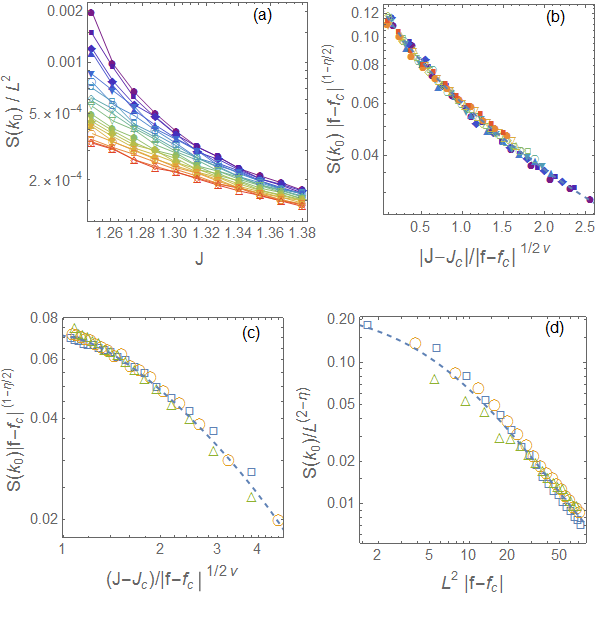}
\caption{ (a) Vortex structure factor peak $S(k_0)$ at different values of $f$, near the dynamical transition for $f = 1/2$.  Temperature $T=0.2$ and system size $L=80$.  $f$ increases from $0.501875$ to   $0.511875$ in $17$ equal steps; (b)  Scaling plot   for $f \ge 0.505$  neglecting finite-size effects, with $J_c=1.24$, $\nu=0.95$ and $\eta=0.7$;
(c) Scaling plot with  $\delta J L^{1/\nu}=c $, a constant ($c= 8.64$),   with  $J_c=1.23$, $ f_c=0.501$,  $\nu=1.0$ and $\eta=0.7$. Open squares correspond to $L=48$, circles to $L=64$ and triangles to $L=80$.  (d) Scaling plot at $J_c=1.23$ with  $ f_c=0.501$ and  $\eta=0.7$. }
\label{sff12}
\end{figure}

In Figs.  \ref{tauf12}a, we show the behavior of the relaxation time $\tau$ obtained from the voltage time correlation function, defined in Eq. (\ref{tcorrel}), as a function of the driving current and  different $f$ near $f=1/2$.  A reasonable data collapse  according to scaling forms of Eq. (\ref{tauscal}) (Figs. \ref{tauf12}b and \ref{tauf12}c) and Eq. (\ref{tauscalb}) (Fig. \ref{tauf12}d) is obtained with $z\approx 2$ and $\nu\approx 1$.

\begin{figure}
\centering
\includegraphics[width=\columnwidth]{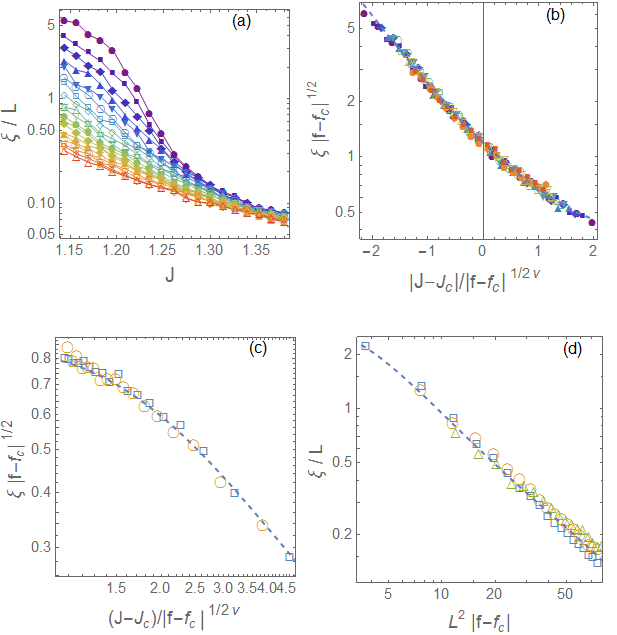}
\caption{ (a) Vortex correlation length $\xi (J,f)$ at different values of $f$, near the dynamical transition for $f = 1/2$.  Temperature $T=0.2$ and system size $L=80$.  $f$ increases from $0.501875$ to   $0.511875$ in $17$ equal steps; (b)  Scaling plot  for $f \ge 0.505$ neglecting finite-size effects, with  $J_c=1.24$, $ f_c=0.5015$ and  $\nu=0.95$;
(c) Scaling plot with  $\delta J L^{1/\nu}=c $, a constant ($c= 8.64$),   with  $J_c=1.20$, $ f_c=0.501$ and  $\nu=1.0$. Open squares correspond to $L=48$, circles to $L=64$ and triangles to $L=80$.  (d) Scaling plot at $J_c=1.23$ with  $ f_c=0.501$. }
\label{xif12}
\end{figure}

The dynamic vortex Mott transition correlates with a change in the structure of the sliding state of the vortex lattice.  As can be seen from the snapshots of the vortex configuration in Fig. \ref{sf}a and Fig.  \ref{sf}b, obtained below and above the critical current $J_c$, respectively, the vortex lattice remains essentially ordered below $J_c$ but has a large number of defects above $J_c$. The transition in the vortex structure can be determined from the behavior of the structure factor $S(\vec k)$, defined in Eq. (\ref{dsf}), which is a measure of the vortex correlations. The sliding ordered state below $J_c$ (Fig. \ref{sf}a)  corresponds to sharp peaks in the structure factor $S(\vec k)$  at the wave vectors $\vec k_o$ of the periodic vortex structure, as expected for a commensurate frustration $f=1/2$.  Above $J_c$ (Fig.  \ref{sf}b), the peaks  broaden and become very small corresponding to the disordered phase. In Figs.  \ref{sff12}a, we show the behavior of the structure factor peak $S(\vec k_0)$  as a function of the driving current $J > J_c$ and  different $f$, near $f=1/2$. As shown in Figs.  \ref{sff12}b, \ref{sff12}c and \ref{sff12}d, a reasonable data collapse  according to the  scaling forms of Eq. (\ref{sfscal}) and Eq. (\ref{sfscalb}) is obtained with $\nu\approx 1$, $\eta \approx 0.7$ and $J_c=1.23$. Finally, in Figs.  \ref{xif12}a, we show the behavior of the vortex correlation length $\xi (J,f)$, defined in Eq. (\ref{cdef}) with ${\bf  k_m} =(2\pi/L) {\bf \hat x}$, as a function of the driving current and  different $f$ near $f=1/2$.  A reasonable data collapse  according to the scaling form of Eq. (\ref{cscal}) (Figs. \ref{xif12}b and \ref{xif12}c) is obtained with $\nu\approx 1$ and  Fig. \ref{xif12}d shows that the data collapse at $J_c$ is consistent with  Eq. (\ref{cscalb}), which does not depend on the critical exponent $\nu$. These estimates of the exponents obtained from vortex correlations agree with the values obtained from the voltage time correlation showing that indeed the dynamic vortex Mott transition for $f=1/2$ corresponds to a structural phase transition of the sliding vortex lattice. 

\begin{figure}
\centering
\includegraphics[width=\columnwidth]{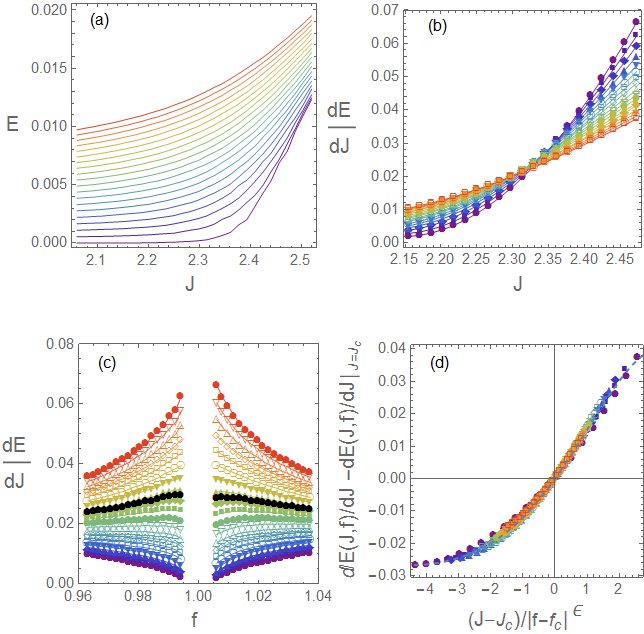}
\caption{(a) Current-voltage characteristics for increasing frustration $f$, near $f=1$. Temperature $T=0.2$ and system size $L=64$. From the bottom up,  $f$ increases from $1$ to $1.03711$ in $19$ equal steps. 
(b) differential resistivity $\frac{d E}{d J}$ as a function of $J$ for $f \ge 1.00586$,  near the transition. (c ) $\frac{d E }{d J}$  as a function of $f$  for different $J$. From the bottom up, $J$ increases from $2.15517 $ to $2.47241 $ in $20$ equal steps. The separatrix $ dE/dJ_{J=J_c} $ is indicated by black dots. (d) Scaling plot of $\frac{d E}{dJ}$  with  $J_c=2.3534$, $ f_c=1.0$ and $\epsilon=0.6$.}
\label{ivf1}
\end{figure}

\subsection{Integer vortex density: $f=1$}

We now consider the dynamic vortex Mott transition near an integer vortex density,  $f=1$, and extract the critical exponents $\epsilon$, $\nu$ and $z$ from the expected scaling behavior as described in Sec. III.  Figure \ref{ivf1}a shows the effect of increasing the frustration $f$ from
$f=1$ on the current-voltage relation at a temperature $T=0.2$. The corresponding behavior of the differential resistivity,  $dE/dJ$,  is shown in  Fig. \ref{ivf1}b. The differential-resistivity curves for different $f$ cross approximately at the same point  $J_c \sim 2.3 $ signaling the dynamic vortex Mott insulator transition. In Fig. \ref{ivf1}c ,  $dE/dJ$ is plotted as a function of $f$ for different currents, where data for both  $f < 1$ and $ f > 1$ are included, showing the reversal of a minimum into a maximum near $f=1$ for increasing currents. The separatrix of the two regimes is indicated by the dotted lines.  In  Fig. \ref{ivf1}d, we plot the data near the transition according to the scaling form of Eq. (\ref{diffscal}), giving an estimate of the critical exponent $\epsilon \approx 0.60(7)$. The  characteristic minimum-maximum reversal of the differential resistivity near $f=1$ for increasing current  and the corresponding data collapse are in good agreement  with the behavior of  the dynamic vortex Mott  transition as observed in the experiments \cite{poccia15,lankhorst2018} with $\epsilon = 2/3$,  supporting the universality of this dynamic transition. 

\begin{figure}
\centering
\includegraphics[width=\columnwidth]{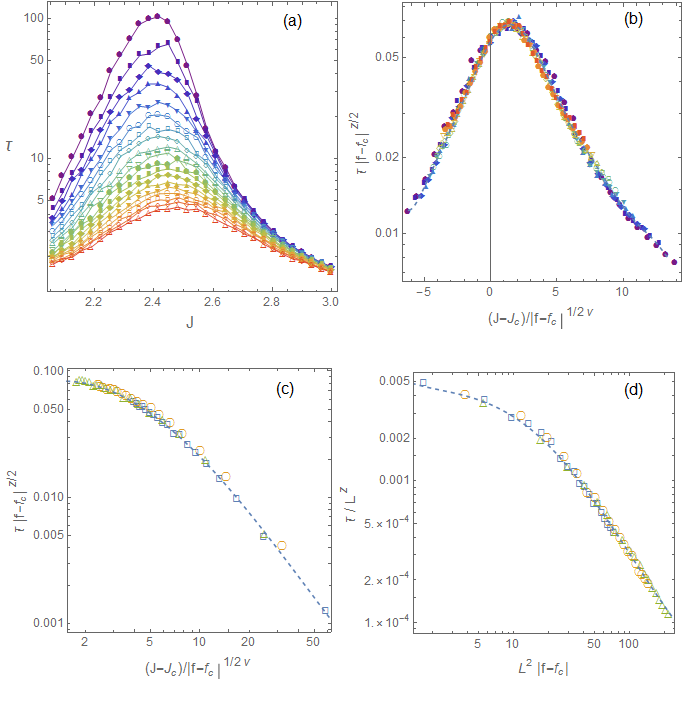}
\caption{ (a) Relaxation time  $\tau(J,f) $  at different values of $f$, near the dynamical transition for $f = 1$.  Temperature $T=0.2$ and system size $L=80$.  $f$ increases from $1.00188$ to   $1.03563$ in $18$ equal steps; (b)  Scaling plot for $f \ge 1.01563$ neglecting finite-size effects with $J_c=2.35$, $ f_c=1.002$, $\nu=0.7$ and $z=2.5$; (c) Scaling plot with  $\delta J L^{1/\nu}=c $, a constant ($c= 85.75$),   with  $J_c=2.36$, $ f_c=1.001$,  $\nu=0.7$ and $z=2.4$. Open squares correspond to $L=48$, circles to $L=64$ and triangles to $L=80$;  (d) Scaling plot at $J_c=2.36$ with $z=2.4$. }
\label{tauf1}
\end{figure}

In Figure \ref{tauf1}a, we show the behavior of the relaxation time $\tau$ obtained from voltage time correlations as a function of the driving current and  different $f$ near $f=1$.  A reasonable data collapse  according to the scaling forms of Eq. (\ref{tauscal}) (Fig. \ref{tauf1}b and \ref{tauf1}c  ) and Eq. (\ref{tauscalb}) (Fig. \ref{tauf1}d)  are obtained with $\nu\approx 0.70(7)$ and $z=2.40(7)$.  The estimated values of $\nu$ and $z$ are indeed compatible with the conjectured values $\nu=3/4$ and $z=7/3$ inferred from the scaling analysis of Sec. IIIA

\section{Conclusions}

We have considered  the dynamic vortex Mott transition in 2D superconducting arrays  in a magnetic field with $f$ flux quantum per plaquette. This nonequilibrium dynamic transition is induced by external driving current and thermal fluctuations  near rational vortex densities 
set by the value of $f$.  Experimentally, the transition has been determined from the scaling behavior of the differential resistivity characterized by a critical exponent $\epsilon$. Recent numerical simulations
of interacting vortex models \cite{eg18,rade17} have demonstrated this behavior only near fractional  $f$. A fine-grid vortex model was introduced, which allowed us to consider both the cases of fractional and integer $f$, and investigate the critical behavior by a scaling analysis and MC simulations. For $f=1/2$, the dynamic transition is accompanied by a structural transition of the sliding vortex lattice. The critical exponents  are consistent with the experimental observations  \cite{poccia15}, and previous numerical results from a standard vortex model \cite{eg18}. The same scaling behavior of the differential resistivity is obtained for  $f=1$, in agreement with experiments \cite{poccia15,lankhorst2018}. However, we find a  correlation-length exponent $\nu \sim 0.75$, which is significantly different from the one expected from mean-field theories, $\nu=1/2$. 
From the scaling analysis we find $\epsilon=1/2 \nu$ and using the experimental results for $\epsilon$ we then conjecture the values $\nu=1$, $z=2$ for  $f=1/2$ and $\nu=3/4$, $z=7/3$ for $f=1$,  which are consistent with the numerical results within the errorbars
Although the critical exponent $\epsilon=2/3$ observed experimentally for integer $f$ can be obtained at the mean-field level by mapping to a  single particle non-Hermitian Hamiltonian \cite{tripathi2016}, our results indicate that the full critical behavior is not described by this approximation and fluctuations should be taken into account. 
%In principle, experiments measuring voltage noise near the transition should allow one to extract the exponents $z$ and $\nu$ and compare with these predictions.
\acknowledgements

%\medskip
The author thanks J. M. Kosterlitz and X. S. Ling for helpul discussions.
This work was supported by  S\~ao Paulo Research Foundation (FAPESP, Grant \#
2018/19586-9), National Council for Scientific and Technological Development-CNPq and computer facilities from CENAPAD-SP.

%\bibliography{finegridbb}

 \newcommand{\noop}[1]{}

\end{document}